\documentclass[letterpaper,twocolumn,10pt]{article}
\usepackage{usenix}

\usepackage{graphicx}
\usepackage{amsmath}
\usepackage{amssymb}
\usepackage{booktabs}
\usepackage{longtable}
\usepackage{multirow}
\usepackage{array}
\usepackage{listings}
\usepackage{pgfplots}
\pgfplotsset{compat=1.18}
\usepackage{pifont}

\newcommand{\codeword}[1]{\texttt{#1}}

\date{}
\title{\Large \bf On the Reliability of LLM-Based Vulnerability
Patching Benchmarks}

\author{
{\rm Dang K Le}\\
Northwestern University\\
dang.le@northwestern.edu
\and
{\rm Wenxuan Shi}\\
Northwestern University\\
wenxuan.shi@northwestern.edu
\and
{\rm Xinyu Xing}\\
Northwestern University\\
xinyu.xing@northwestern.edu
}

\begin{document}

\maketitle

\begin{abstract}
  Large language models (LLMs) have shown remarkable potential for
  automated vulnerability patching, with benchmarks such as SWE-bench
  driving rapid progress. However, through extensive experience
  developing, running, and stress-testing such evaluation frameworks,
  we identify critical but under-examined pitfalls in current
  benchmarking practices that can substantially distort reported
  results. These pitfalls span three dimensions: (1)
  \emph{agent-level} factors, where prompting strategies, tool
  availability, and detailed instructions inflate success rates
  noticeably without improving developer-aligned patch quality; (2)
  \emph{framework-level} factors, where permission misconfigurations,
  infrastructural bugs, and timeout issues silently degrade or
  artificially boost agent performance; and (3) \emph{dataset-level}
  factors, where reliance on bug reports and single proof-of-concept
  (PoC) tests fails to capture whether patches address root causes or
  adhere to developer design intent. To demonstrate the effects of
  these shortcomings, we curate a dataset of 112 real-world historical
  bugs across 84 open-source C/C++, Go, and Rust projects, each equipped
  with PoC tests, regression tests, and added \emph{developer's tests}
  that evaluate whether generated patches align with the original
  developers' design principles. Through three controlled experiments
  and multiple case studies, we demonstrate that while LLMs can achieve
  high PoC passing rates in ideal scenarios, many factors in executing
  a benchmark can noticeably lower the success rate. Moreover, their
  developer's test passing rates remain low and improve only marginally
  with newer models, revealing that models are increasingly adept at
  suppressing symptoms but not at producing patches that satisfy
  upstream developer-authored tests. Our results show that current
  benchmark scores are highly sensitive to evaluation design choices,
  and we provide practical guidelines for making future evaluations
  more rigorous, reliable, and reproducible.
\end{abstract}

\section{Introduction}
\label{sec:intro}

Automated program repair (APR) with large language models (LLMs) has
progressed from synthetic code-completion
tasks~\cite{ribeiro_large_2023} to end-to-end agents that localize,
patch, and validate fixes against real-world software
projects~\cite{xia_automated_2023,zhang_systematic_2026}.
Benchmarks such as SWE-bench~\cite{jimenez_swe-bench_2024} and its
multi-language
successors~\cite{zan_multi-swe-bench_2025,mei2024arvoatlasreproduciblevulnerabilities,hwiwon2025sec}
have become the de facto yardstick~\cite{martinez_whats_2026}: a new
model is judged by the fraction of issues it
``resolves,''~\cite{liu_critical_2021} and leaderboards are updated weekly.
This enthusiasm has spawned a vibrant ecosystem of agent frameworks,
fine-tuned models, and training datasets derived from benchmark
traces~\cite{yang_survey_2025,machacek_impact_2025}.

Unfortunately, the same ecosystem is remarkably opaque about
\emph{how} those numbers are produced.
A typical benchmark run places an agent in a buggy repository with a
bug report, exposes some set of tools and auxiliary artifacts, and
scores the resulting patch against one or more automated oracles.
Seemingly small choices, such as whether the harness tells the agent
how to build the project, or whether the target's git history is
included in the workspace, can significantly affect the reported pass rate.
Worse, an agent can ``pass'' a benchmark case by silencing the
symptom used to score it while leaving the underlying defect, or even
the whole codebase, demonstrably worse off~\cite{wang_are_2025}.
If these results are then used to train the next generation of code
models, the pitfalls are baked into model weights~\cite{xu_benchmark_2024}.

In this paper we argue that the \emph{benchmarking pipeline itself},
not just the model or the dataset, is a first-class research artifact
that deserves principled design and critical scrutiny:
\emph{benchmark scores are highly sensitive to evaluation design and
can materially misrepresent model capability.}
Drawing on our experience developing and stress-testing LLM patching
agent harnesses across three languages, we identify a taxonomy of
pitfalls along three dimensions:
\textbf{Agent-level factors} (\S\ref{sec:factors:agent}): prompting
choices, tool permissions, feedback oracles, and auxiliary capabilities
such as Internet access; \textbf{framework-level factors}
(\S\ref{sec:factors:framework}): permission misconfigurations,
timeouts, hidden agent mechanics, and post-agent test interference; and
\textbf{dataset-level factors} (\S\ref{sec:factors:dataset}): how
vulnerabilities are described, whether PoCs are provided, how the
source is packaged, and how ``success'' is operationalized.

To study these factors empirically, we curate 112 historical bugs
across 84 C/C++, Go, and Rust projects, each packaged in a
reproducible Docker harness (\S\ref{sec:method}) with PoCs,
regression tests, and developer's tests
(\S\ref{sec:method:devtest})---tests that encode the developer's
design intent, not just the absence of a crash.
We then run three controlled experiments and five case studies to
probe how each factor shifts apparent patch quality
(\S\ref{sec:major-experiments}, \S\ref{sec:case-studies}).

Our experiments are designed to isolate \emph{principles} rather than
individual knobs.
For example, across a modern frontier model, we observe that detailed
build and test instructions lift PoC-pass rates by a large margin in
C/C++ and much less so in Go/Rust---because in Go/Rust the build
system is uniform and discoverable from the repository, whereas C/C++
projects' build systems are heterogeneous.
This illustrates one of the principles we demonstrate in this paper:
\emph{the reported success rate of a benchmark is only meaningful
relative to the information its harness makes available to the agent}.
We generalize this style of observation into a set of guidelines
(\S\ref{sec:guidelines}) for how benchmarks should be built and reported.

\noindent \textbf{Contributions.} We make the following contributions:
\textbf{(1)} A three-level taxonomy of pitfalls in LLM vulnerability
patching benchmarks, grounded in concrete failures we observed in practice.
\textbf{(2)} A controlled evaluation dataset and harness designed to
isolate benchmarking factors: 112 real-world security-relevant bugs
across three languages, each packaged with reproducible builds, PoCs,
regression tests, and developer's tests derived from upstream commits.
\textbf{(3)} Three controlled experiments quantifying how prompting,
tool availability, PoC access, and model choice move both PoC-pass and
developer-test-pass rates; and five case studies that show how
``incidental'' framework issues can silently dominate headline numbers.
\textbf{(4)} A survey of 13 existing real-world patching benchmarks and
vulnerability datasets along the same factor axes, showing that most
exhibit at least one of the pitfalls we characterize.
\textbf{(5)} A set of concrete guidelines and a reporting checklist for
future benchmark designers.

\noindent \textbf{Scope. This work is \emph{not} intended to propose another benchmark for
model ranking}. Instead, the dataset is deliberately designed to
highlight evaluation pitfalls under controlled conditions rather than
to support competitive comparison. Our contribution is
methodological: to strengthen the validity, interpretability, and
reproducibility of reported APR results and of the training signals
subsequently derived from them.

\section{Background and Motivation}
\label{sec:background}

\subsection{LLM-Based Vulnerability Patching}
\label{sec:background:llm}

Modern LLM patching systems typically wrap a frontier model (e.g.,
Claude, GPT, Gemini) in an \emph{agent harness}: a scaffold that
exposes shell, file-editing, and sometimes web-browsing tools, loops
over model turns, and terminates when the agent emits a ``done''
signal or hits a step/turn
budget~\cite{yang_swe-agent_2024,zhang_autocoderover_2024}.
A typical patching task gives the agent: (a) a textual description of
the bug (often a GitHub issue), (b) access to the target repository
at a buggy commit, and (c) optionally a PoC input and/or instructions
related to building and testing the
target~\cite{yu_patchagent_2025,zhang_autocoderover_2024}.
The agent is graded by an automated oracle, usually whether the PoC
no longer reproduces the crash and whether a reference test suite still passes.

\subsection{``Just Another Benchmark'' Is Not Enough}
\label{sec:background:gap}

Existing benchmarks occupy a wide design space, but their evaluation
protocols have converged on a surprisingly narrow notion of
correctness: \emph{did the supplied test(s)
pass?}~\cite{yu_patchagent_2025,zhang_fixing_2024}
This definition is seductive because it is cheap and objective.
But it conflates very different outcomes:
\textbf{(1)} the agent produced a patch the original maintainer would
have accepted; \textbf{(2)} the agent produced a patch that makes the
oracle happy, e.g., by short-circuiting the crashing code path,
disabling the feature, or special-casing the PoC input; or
\textbf{(3)} the agent produced a patch that looks perfect, by taking
advantage of a pitfall in the benchmark, such as git history, Internet
access, permission misconfiguration, etc.
All of the outcomes above appear identical on a leaderboard.
We argue that the difference matters for two downstream uses:

\noindent \textbf{Deployment.} If an APR system is proposed as a
developer tool or as part of an autonomous pipeline, a fix that
silences the PoC but breaks a corner of the feature is strictly worse
than no fix at all, because it provides false assurance and masks the
underlying vulnerability.

\noindent \textbf{Training.} Increasingly, agent trajectories that
``pass'' the benchmark are used as supervised or
reinforcement-learning
signal~\cite{zhou_learning_2026,luo_agent_2025,tian_learning_2023}.
Every false-positive pass is a data point that teaches the model
to produce similar surface-level patch.

\section{A Taxonomy of Benchmarking Pitfalls}
\label{sec:factors}

We organize the pitfalls into three levels, corresponding to the
three artifacts a benchmark designer controls: the \emph{agent} (what
the model is allowed to do), the \emph{framework} (how its behavior
is measured), and the \emph{dataset} (what it is measured on).
We define a \emph{factor} as any variable in the evaluation pipeline
that can change the measured success rate without changing the underlying model.
We refer to factors studied via controlled ablations as
\emph{systematic}, and those revealed through individual failures as
\emph{incidental}; both can substantially affect reported performance.
These factors are not independent. In practice, their interactions
can amplify effects. Throughout this section and in
\S\ref{sec:major-experiments}, \S\ref{sec:case-studies}, we analyze
factors in isolation for clarity, but we note that real-world
benchmarks might exhibit multiple interacting factors simultaneously
(\S\ref{sec:survey}).

\subsection{Agent-Level Factors}
\label{sec:factors:agent}

\noindent \textbf{A1. Build and test instructions in the prompt
(systematic).} An impactful prompt-level variable we found is whether
the agent is
told, in-context, how to \emph{build} and \emph{test} the project.
Without such guidance, a capable agent must infer the build system
from files scattered across the repository; with it, the agent can
validate its own changes after every edit.
We show in \S\ref{sec:exp:buildtest} that supplying these
instructions raises PoC-pass rates by $\approx20$\% in C/C++ cases
and by a far smaller margin in Go/Rust cases, where the build system
is more uniform.

\noindent \textbf{A2. Internet and web-search access (incidental).}
LLM agents are increasingly given live web-search and URL-fetch
tools~\cite{soni_coding_2025}.
For a benchmark drawn from public GitHub issues, these tools leak
ground truth in several ways.
We observed several case traces where the agent directly searched for
the issue number and retrieved the fix verbatim (\S\ref{sec:case:websearch}).


\subsection{Framework-Level Factors}
\label{sec:factors:framework}

\noindent \textbf{F1. Permission misconfiguration (incidental).} A
harness typically mounts the target source tree together with
sensitive metadata related to the issue.
If the agent is granted overly broad permissions, it may access these
files and extract information about the ground-truth fix.
Conversely, if permissions are too restrictive (e.g., preventing
installation of dependencies required for a legitimate build), the
agent may underperform for reasons unrelated to its capabilities.

\noindent \textbf{F2. Timeouts and hidden mechanics (incidental).}
Closed-source agent systems contain behaviors that are not obvious
from the documentation alone.
Evidently, we observed one such behavior in Claude Code, leading to
unintended timeouts (\S\ref{sec:case:haiku}).
Analogously, slow-building projects (e.g., large C/C++ code bases) can exhaust wall-clock budgets before the agent has
had a chance to test.
These effects produce false negatives that are indistinguishable from
real model failures.

\noindent \textbf{F3. Agent-written tests interfering with the
evaluation pipeline (incidental).} Some agents routinely create their
own tests to validate a fix.
If the evaluation framework subsequently applies additional tests to
a file that the agent has already modified to add its own tests, the
test application step may fail due to conflicts.
As a result, the case is marked as a failure even though the patch
itself is correct (\S\ref{sec:case:testinterference}).
This is particularly problematic: the agent follows the intended
workflow and produces a valid fix, yet the failure arises entirely
from the evaluation pipeline rather than the agent's behavior.

\subsection{Dataset-Level Factors}
\label{sec:factors:dataset}

\noindent \textbf{D1. Oracle granularity: PoC versus developer's test
(systematic).} The single most common oracle in APR benchmarks is ``does the PoC
still crash?''~\cite{gao_beyond_2021,xu_vfix_2019,zhang_fixing_2024}
This is a necessary but insufficient condition for correctness.
We propose \emph{developer's tests} (\S\ref{sec:method:devtest}) as a
practical middle ground between PoC-only and full regression testing:
a targeted, author-authored oracle that encodes the fix's \emph{intent}.

\noindent \textbf{D2. PoC format and framing (systematic).} When a PoC
is included, how it is presented matters.
A raw binary input tells the agent nothing about intent, while a test
harness (e.g., a Go test
function) reveals which API is exercised and with what expectations.
We show in \S\ref{sec:exp:poc} that PoCs presented as source-level
harnesses yield markedly higher PoC-pass rates than binary PoCs,
independent of patch quality.

\noindent \textbf{D3. Source provenance and contamination
(incidental).} The method used to retrieve the source code determines what
incidental signals are exposed in the workspace.
Shallow clones, tarballs, and git worktrees each reveal different
amounts of history.
In particular, the presence of an unsanitized Git history is risky: a
future commit in the history may contain the upstream fix
for the buggy version (\S\ref{sec:case:gitlog}).

\noindent \textbf{D4. Language balance (systematic).} Most benchmarks
are dominated by a single language (typically Python
or Java for SWE-bench variants).
We find that the sensitivity of patch quality to specific factors
varies sharply by language: C/C++ benchmarks are more sensitive to
harness quality, Go/Rust benchmarks are more sensitive to PoC framing.

\section{Methodology and Dataset}
\label{sec:method}

We built a reproducible evaluation infrastructure in which each of
the above factors can be toggled in isolation.
This section describes the dataset, the harness, and the agent/model
choices used throughout \S\ref{sec:major-experiments}--\S\ref{sec:case-studies}.

\subsection{Dataset}
\label{sec:method:dataset}

Our dataset contains 112 real-world historical bugs spanning 84
open-source projects across three languages, drawn from public GitHub
issues, GitHub Security Advisories (GHSA), RUSTSEC advisories, and
curated commit histories.
The per-language breakdown is summarized in
Table~\ref{tab:dataset:overview}; the full dataset instances and
their metadata are listed in Appendix~\ref{app:dataset}.

\begin{table}[h]
  \centering
  \caption{Per-language dataset breakdown. ``Bug-desc.\ source'' is the primary
    artifact provided to the agent as the bug description; ``PoC form'' is the
  format of the proof-of-concept input bundled with each case.}
  \label{tab:dataset:overview}
  \renewcommand{\arraystretch}{1.15}
  \footnotesize
  \begin{tabular}{@{}lccp{2.0cm}p{1.9cm}@{}}
    \toprule
    \textbf{Language} & \textbf{Proj.} & \textbf{Cases} &
    \textbf{Bug-desc.\ source} & \textbf{PoC form} \\
    \midrule
    C/C++ & 10 & 25 & GitHub/GitLab issue + ASan trace & binary/text
    blob \& C source harness \\
    Go    & 51 & 62 & GH issue, GHSA   & Go source harness \\
    Rust  & 23 & 25 & GH issue, RUSTSEC, GHSA & Rust
    source harness \\
    \midrule
    Total & 84 & 112 & --- & --- \\
    \bottomrule
  \end{tabular}
\end{table}

The C/C++ slice spans 10 projects: interpreters,
image/data-processing libraries, command-line tools, and binary tooling.
The Go slice spans 51 projects, including web frameworks,
infrastructure tooling, and security-relevant libraries.
The Rust slice spans 23 projects with memory-safety and logic
CVEs, including RUSTSEC advisories affecting data structure \& cryptographic
libraries, and HTML sanitizers.

Each case is pinned to a specific \emph{trigger commit} and
\emph{patch commit}, and is packaged as a self-contained Docker
build. The per-case directory includes:
\textbf{(1)} target-specific scripts for building, reproducing the PoC,
and running regression tests. The PoC reproduction script is
carefully designed to return a non-zero exit code \emph{only}
when a crash associated with the vulnerability is triggered,
rather than when the program itself exits with a non-zero status
(e.g., due to invalid inputs).
\textbf{(2)} PoC inputs, which are binary artifacts (most C/C++ cases) or
source-level harnesses (all Go and Rust cases).
\textbf{(3)} The upstream developer's patch, used exclusively by the
ground-truth validator (\S\ref{sec:method:oracle}) and never
exposed to the agent.
\textbf{(4)} The \emph{developer's test} derived from the upstream commit
(\S\ref{sec:method:devtest}).
\textbf{(5)} The bug description provided to the agent: a textual
issue/advisory write-up, a sanitizer crash trace, or both,
depending on what is available upstream.
\textbf{(6)} A YAML configuration file that captures metadata and pins all
of the above components.

\noindent \textbf{Data collection and curation.} Cases were collected
by scraping historical security issues and fix
commits, restricted to projects whose build-and-test cycle completes
in under half an hour in our containerized harness.
For C/C++ cases, PoCs are the binary inputs attached to the upstream
issue; the bug description is the GitHub or GitLab issue text
combined with a sanitizer crash trace we reproduce in the harness.
For Go cases, the bug description is taken from the upstream issue or
GHSA advisory; PoCs are source-level Go programs, of which 25 were
imported from the upstream issue or advisory and 37 were minimal
harnesses we hand-wrote when no upstream PoC was available.
For Rust cases, the bug description is taken from RUSTSEC advisories,
GHSA advisories, or the upstream issue; PoCs are source-level Rust
programs sourced from the RUSTSEC advisory's reference exploit when
available, with the remaining cases using harnesses adapted from the
upstream report or hand-written by us.
Each case is validated by our ground-truth validator
(\S\ref{sec:method:oracle}): with the upstream patch applied, the
target must build successfully, avoid crashing on the PoC, and pass
all regression and developer's tests.
A case is admitted into the dataset only if it passes all four
phases; cases that failed any phase were either repaired or
discarded, yielding the final 112 curated instances.

\subsection{Developer's Tests}
\label{sec:method:devtest}

A \emph{developer's test} is a unit or integration test authored by
the developers and extracted from the upstream fix commit. In mature
projects, when developers fix an issue, they often introduce
additional tests to validate one or more of the
following~\cite{mi_empirical_2016}:
\textbf{(1)} the patch addresses the root cause rather than simply its
symptoms; \textbf{(2)} the patch preserves the intended functionality
of the modified code, especially in cases where existing regression
tests do not adequately cover the affected area; or \textbf{(3)} the
patch conforms to the software's specifications, coding conventions,
and intended design.
In our dataset, we extract these tests from the upstream fix commit
and reconstruct them as a patch named \codeword{test.diff}, which can
be applied to the pre-fix source tree. This allows us to execute
developer's tests in the same manner as the project's regression suite.
Conceptually, a developer's test answers a different question than a PoC:
\begin{center}
  \footnotesize
  \renewcommand{\arraystretch}{1.1}
  \begin{tabular}{@{}lp{0.58\columnwidth}@{}}
    \toprule
    \textbf{Oracle} & \textbf{Question answered} \\
    \midrule
    Build & Is the codebase still syntactically/linkable valid? \\
    PoC & Does the known crashing input still crash? \\
    Regression & Do pre-existing regression tests still pass? \\
    Developer's test & Does the patch behave as the developer
    intended on inputs adjacent to the bug? \\
    \bottomrule
  \end{tabular}
\end{center}

\noindent \textbf{Example: PHP \codeword{range()} type confusion (GH-13094).}

To illustrate the difference between PoC-based evaluation and
developer-intent evaluation, consider PHP issue
GH-13094~\cite{php_issue_13094},
a type-confusion bug in \codeword{range()}.
The bug arises because mixed-type inputs can be misclassified as
string inputs, leading the function to dereference a non-string
operand and crash on the PoC \codeword{range(9.9, [])}.

At a high level, the vulnerable code uses an arithmetic predicate
over internal type tags to decide whether to treat the arguments as strings:

The relevant code, simplified for presentation, is the following:

\begin{lstlisting}[language=C]
if (start_type >= IS_STRING || end_type >= IS_STRING) {
    if (start_type + end_type < 2 * IS_STRING) {
        goto handle_numeric_inputs;
    }
    /* unsafe string access */
}
\end{lstlisting}

Two patches can plausibly “fix” this bug.
The upstream developer patch replaces the brittle arithmetic
condition with a direct semantic check for mixed non-string inputs,
preserving PHP's intended coercion behavior.
A second, more defensive patch instead adds type checks immediately
before the unsafe string dereference and throws an error when either
operand is not a string.
Both patches suppress the original crash and pass the pre-existing
PHP regression suite.

However, the two patches are not behaviorally equivalent.
PHP's \codeword{range()} is intended to permit mixed numeric/string
inputs~\cite{php_range_manual} such as
\codeword{range(9.9, \textquotesingle 0\textquotesingle)}, coercing
the string operand and producing a numeric range.
The upstream fix preserves this behavior; the defensive patch rejects
it with \codeword{ValueError}.
The developer's test added in the upstream commit encodes exactly
this expected behavior by checking the output of \codeword{range(9.9,
\textquotesingle 0\textquotesingle)} and asserting it to an exact array value.
As a result, the upstream patch passes the developer's test, while
the defensive patch fails.

\noindent \textbf{What the example shows.} PHP GH-13094 illustrates
that a benchmark which scores patches with
the PoC and the pre-patch regression suite alone treats the
developer's and defensive patches as identical successes, while a
benchmark that adds the developer's test treats only the developer's
patch as a success.
This is the kind of distinction we want LLM patching benchmarks to be
capable of drawing.

\noindent \textbf{Authoring fidelity.} Because developer's tests are
derived directly from upstream commits,
they reflect the maintainer's notion of correctness rather than ours.
We only normalize test infrastructure (e.g., expected-output
formatting, harness wiring) when necessary to run in our sandbox.
Cases whose upstream commit does not add or meaningfully modify a
test are excluded from the dataset rather than instrumented post-hoc
by us; we discuss the resulting threats in \S\ref{sec:threats}.

\subsection{Oracle Pipeline}
\label{sec:method:oracle}

After the agent produces a patch, we evaluate it in four phases:
\textbf{(1) Build:} re-build the target project, with the LLM-generated
patch applied, to completion. \textbf{(2) PoC:} execute each PoC ten
times and require that no run result in a vulnerability-related crash.
Repeating executions helps capture vulnerabilities with
non-deterministic trigger rates (e.g., race conditions), ensuring they
are reliably detected. \textbf{(3) Regression:} run the target
project's built-in regression test (before applying the developer's
test). \textbf{(4) Developer's test:} apply the developer's test patch
file to the patched tree and re-run the regression test, which now
includes the added developer's test.

A case is considered \emph{fully resolved} only if all four phases pass.
We separately report the \emph{PoC pass rate}, \emph{the regression
pass rate}, and the \emph{developer's test pass rate} throughout
\S\ref{sec:major-experiments}.

\subsection{Agents and Models}
\label{sec:method:agents}

We use two distinct agent/model combinations, matched to the scale of
each experiment:

\noindent \textbf{Controlled experiments
(\S\ref{sec:major-experiments}).} We use
\emph{OpenCode}~\cite{opencode} in combination with
\emph{Claude Sonnet 4.0}~\cite{claude_models}.
OpenCode is open-source, which allows us to audit its behavior
end-to-end, and Sonnet 4.0 offers a cost/quality tradeoff that makes
full-factor sweeps across hundreds of runs tractable.

\noindent \textbf{Case studies (\S\ref{sec:case-studies}).} We use
\emph{Claude Code}~\cite{claude_code} in combination
with \emph{Claude Sonnet 4.5}. Claude Code is a production
closed-source agent with non-trivial internal mechanics, making it
useful for exposing framework-level pitfalls that are difficult to
observe in simpler harnesses.

Across both settings, the evaluation harness is held fixed: the agent
operates on the checked-out buggy repository inside a containerized
environment, while prompt components such as build/test instructions
and PoC exposure are toggled through configuration so that individual
factors can be isolated cleanly. Model access is routed through a
LiteLLM~\cite{litellm}
proxy for uniform invocation and logging. Appendix~\ref{app:infra}
summarizes the container layout, isolation model, runtime controls,
and evaluation pipeline; full implementation details are included in
the artifact.

\section{Experiments of Systematic Factors}
\label{sec:major-experiments}

We run three controlled experiments that hold the dataset constant
and vary one agent- or dataset-level factor at a time. All
experiments are performed under restrained container environments,
hosted in typical Linux environment with 16 CPU cores, 32GB RAM,
under normal workloads. The artifact includes per-case phase-level
results for all conditions.

\subsection{Experiment 1: Build/Test Instructions}
\label{sec:exp:buildtest}

\noindent \textbf{Setup.} We run the full 112-case dataset twice with
OpenCode/Claude~Sonnet~4.0.
In one condition, the prompt includes the verbatim contents of the
scripts the evaluator uses to build the project and run regression
test; in the other, both are omitted, forcing the agent to infer how
to build and test the project from the repository itself.
All other conditions are fixed.

\noindent \textbf{Expected effect.} \emph{Providing build and test
  instructions should primarily improve PoC-pass rates by making
  self-validation easier, with the largest effect in C/C++ projects
  whose build systems are heterogeneous and less discoverable than
those of Go and Rust.}

\noindent \textbf{Results.} Table~\ref{tab:exp:buildtest} reports
per-language pass rates for both conditions across three test oracle phases.

\begin{table}[h]
  \centering
  \caption{Experiment 1: PoC, regression, and developer-test pass
    rates with and without build/test instructions in the prompt,
    broken down by language. Values are percentages computed over
  $\textrm{pass}+\textrm{fail}$ outcomes (denominator = total cases per row).}
  \label{tab:exp:buildtest}
  \renewcommand{\arraystretch}{1.15}
  \footnotesize
  \begin{tabular}{@{}lcccccc@{}}
    \toprule
    & \multicolumn{3}{c}{\textbf{With instructions (Exp.~1)}} &
    \multicolumn{3}{c}{\textbf{Without (Exp.~2)}} \\
    \cmidrule(lr){2-4}\cmidrule(lr){5-7}
    \textbf{Lang.} & PoC & Reg. & Dev. & PoC & Reg. & Dev. \\
    \midrule
    C/C++           & 92.0 & 88.0 & 36.0 & 72.0 & 92.0 & 28.0 \\
    Go              & 69.4 & 100.0 & 41.9 & 72.6 & 100.0 & 40.3 \\
    Rust            & 84.0 & 100.0 & 56.0 & 80.0 & 96.0 & 56.0 \\
    \midrule
    Overall         & 77.7 & 97.3 & 43.8 & 74.1 & 97.3 & 41.1 \\
    \bottomrule
  \end{tabular}
\end{table}

\noindent \textbf{Interpretation.}
Including build/test instructions raises overall PoC-pass from 74.1\%
to 77.7\%, but the effect is highly uneven across languages. The
C/C++ slice improves by 20 percentage points (72.0\% to 92.0\%),
whereas Rust changes only modestly (80.0\% to 84.0\%) and Go is
nearly unchanged (72.6\% to 69.4\%, a slight negative within the
noise of a single run). This matches the structure of the ecosystems
in our dataset: Go and Rust projects usually expose recoverable
defaults for building and testing such as \codeword{go build} \&
\codeword{go test} and \codeword{cargo build} \& \codeword{cargo
test}, while C/C++ projects rely on far more heterogeneous build and
test procedures (such as CMake, Autotools, hand-written Makefiles,
project-specific shell wrappers, etc.). By contrast, regression-pass
rates remain constant overall (97.3\% in both conditions), and
developer-test-pass changes only slightly (41.1\% to 43.8\%).
Together, these results suggest that build/test instructions mainly
alter the benchmark's information budget---especially in
C/C++---rather than patch quality as measured by our stronger
developer-test oracle.

\noindent \textbf{Takeaway.}
The headline PoC-pass rate of an otherwise identical benchmark can
shift by 20 percentage points in C/C++ purely as a function of
whether the harness discloses build/test instructions in the prompt.
Because current benchmark papers disclose this detail inconsistently,
reported scores across benchmarks are not directly comparable, and
the magnitude of the discrepancy is itself language-dependent.

\subsection{Experiment 2: PoC Framing}
\label{sec:exp:poc}

\noindent \textbf{Setup.} We study two subsets of the dataset
according to the natural form of their PoC artifacts. The Go subset
(62 cases) uses source-harness PoCs written in Go that exercise
specific APIs in the target package and trigger the vulnerability;
the C/C++ subset (25 cases) uses binary or text-blob PoCs fed
directly to an executable. For each subset, we compare a no-PoC
condition against one in which the PoC is exposed to the agent:
inline source code for Go and a file path for C/C++. Build and test
instructions are included in both conditions.

\noindent \textbf{Expected effect.} \emph{Exposing a PoC should help
  most when the artifact is readable and reveals API-level intent, as
  with source-level harnesses. Opaque binary PoCs should offer much
  less benefit because they indicate only that some input crashes the
program, not why or at what semantic level.}

\noindent \textbf{Results.} Table~\ref{tab:exp:poc} shows a strong
contrast between these two PoC forms.

\begin{table}[h]
  \centering
  \caption{Experiment 2: effect of PoC form on PoC, regression, and
    developer-test pass rates. Each row is restricted to the language
    subset in which the corresponding PoC form is natural; the no-PoC
  rows are the language-matched baselines from Experiment~1.}
  \label{tab:exp:poc}
  \renewcommand{\arraystretch}{1.15}
  \footnotesize
  \begin{tabular}{@{}lccc@{}}
    \toprule
    \textbf{PoC form} & \textbf{PoC} & \textbf{Reg.} & \textbf{Dev.} \\
    \midrule
    No PoC (Go, $n{=}62$)              & 69.4 & 100.0 & 41.9 \\
    Source-harness PoC (Go, $n{=}62$)  & 95.2 & 98.4  & 48.4 \\
    \midrule
    No PoC (C/C++, $n{=}25$)           & 92.0 & 88.0  & 36.0 \\
    Binary blob PoC (C/C++, $n{=}25$)  & 92.0 & 92.0  & 36.0 \\
    \bottomrule
  \end{tabular}
\end{table}

\noindent \textbf{Interpretation.}
In Go, adding the source-harness PoC raises PoC-pass from 69.4\% to
95.2\% and developer-test-pass from 41.9\% to 48.4\%. In C/C++,
exposing the binary blob yields no measurable change in PoC-pass
(92.0\% in both conditions) and no change in developer-test-pass
(36.0\% in both), with only a small increase in regression-pass
(88.0\% to 92.0\%). The asymmetry is consistent with the information
each PoC carries. A source-level Go harness exposes both the crashing
call site and the API-level expectation, whereas a binary input is
largely opaque. In our C/C++ cases, that opacity matters even more
because the bug description already includes an ASan stack trace that
localizes the fault.

\noindent \textbf{Takeaway.}
The benefit of PoC access depends much more on its representation
than on the mere fact that a PoC is provided.
Two benchmarks that nominally ``provide a PoC'' are therefore not
comparable unless they also disclose the PoC's representation.
Crucially, the gains accrue far more to oracle alignment than to
developer-aligned patch quality.

\subsection{Experiment 3: Model Generations}
\label{sec:exp:models}

\noindent \textbf{Setup.} We run the full 112-case dataset three
times with the same OpenCode harness used in Experiment 1 (build/test
instructions included, no PoC body), varying only the underlying
model, from older baseline to current frontier reasoning model:
Claude Sonnet 4.0, Claude Sonnet 4.5, and Claude Opus 4.7. We record
overall PoC-, regression-, and developer-test-pass rates across all
three languages.

\noindent \textbf{Expected effect.} \emph{As models improve, PoC-pass
  should rise faster than developer-test-pass. If so, PoC-only
  evaluation will increasingly overstate how close the benchmark is to
being solved.}

\begin{figure}[h]
  \centering
  \begin{tikzpicture}
    \begin{axis}[
        width=0.95\columnwidth,
        height=3.3cm,
        xlabel={Model generation},
        ylabel={Pass rate (\%)},
        symbolic x coords={Sonnet 4.0, Sonnet 4.5, Opus 4.7},
        xtick=data,
        ymin=30, ymax=105,
        ytick={40,60,80,100},
        grid=both,
        grid style={gray!20},
        legend style={
          font=\footnotesize,
          at={(0.5,-0.22)},
          anchor=north,
          draw=none
        },
        legend columns=3,
        legend cell align=left,
        mark size=2pt,
        every axis plot/.append style={very thick},
        tick label style={font=\footnotesize},
        label style={font=\footnotesize},
      ]
      \addplot[mark=*,blue!70!black]
      coordinates {(Sonnet 4.0,77.7) (Sonnet 4.5,84.8) (Opus 4.7,90.2)};
      \addlegendentry{PoC pass}

      \addplot[mark=square*,red!70!black]
      coordinates {(Sonnet 4.0,97.3) (Sonnet 4.5,94.6) (Opus 4.7,94.6)};
      \addlegendentry{Regression pass}

      \addplot[mark=triangle*,dashed,green!50!black]
      coordinates {(Sonnet 4.0,43.8) (Sonnet 4.5,48.2) (Opus 4.7,60.7)};
      \addlegendentry{Developer-test pass}
    \end{axis}
  \end{tikzpicture}
  \caption{Experiment 3: PoC-pass, regression-pass, and
    developer-test-pass rates across three Claude generations (overall,
    $n{=}112$). PoC-pass climbs steadily toward saturation;
  developer-test-pass improves but stays well below; regression-pass is flat.}
  \label{fig:exp:models}
\end{figure}
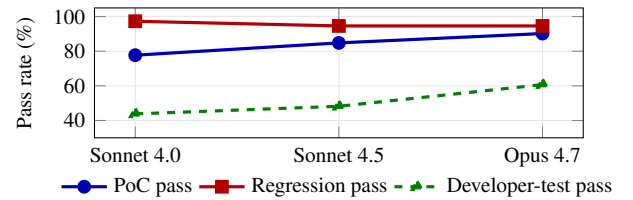

\noindent \textbf{Results.} Figure~\ref{fig:exp:models} shows the
three pass rates as a function of model generation; the underlying
overall numbers are summarized in Table~\ref{tab:exp:models}.

\begin{table}[h]
  \centering
  \caption{Experiment 3: overall pass rates (\%) across Claude model
    generations and the persistent PoC$-$Dev gap. Within parentheses we
    report the share of PoC-passing cases that nonetheless fail the
  developer's test.}
  \label{tab:exp:models}
  \renewcommand{\arraystretch}{1.15}
  \footnotesize
  \begin{tabular}{@{}lcccc@{}}
    \toprule
    \textbf{Model} & \textbf{PoC} & \textbf{Reg.} & \textbf{Dev.} &
    \textbf{PoC$-$Dev (PoC-only \%)} \\
    \midrule
    Sonnet 4.0 & 77.7 & 97.3 & 43.8 & 33.9~~(43.7\%) \\
    Sonnet 4.5 & 84.8 & 94.6 & 48.2 & 36.6~~(43.2\%) \\
    Opus 4.7   & 90.2 & 94.6 & 60.7 & 29.5~~(32.7\%) \\
    \bottomrule
  \end{tabular}
\end{table}

\noindent \textbf{Interpretation.}
PoC-pass rises steadily across model generations, from 77.7\% to
84.8\% to 90.2\%, approaching saturation by Opus 4.7.
Developer-test-pass also improves, but much more slowly: 43.8\% to
48.2\% to 60.7\%. The resulting PoC--Dev gap remains large throughout,
shrinking only from 33.9 to 29.5 percentage points, and even with
Opus 4.7, 32.7\% of PoC-passing cases still fail the developer's
test. In other words, roughly one in three patches that appear
successful under a PoC-only oracle would still be rejected by an
oracle aligned with maintainer intent. Regression-pass remains nearly
flat (97.3\%, 94.6\%, 94.6\%), suggesting that pre-existing
regression suites do little to distinguish symptom suppression from
developer-acceptable repair.

\noindent \textbf{Takeaway.}
These results show that PoC-pass is a misleading progress signal near
the top of the capability curve: a PoC-only benchmark would suggest
that the task is nearly solved, whereas developer's tests reveal
substantial remaining headroom.
Every benchmark currently using only PoC/test oracles is therefore
reporting a scaling curve that overstates how much of the remaining
headroom matters to practitioners.

\subsection{Cross-Cutting Observations}
\label{sec:exp:cross}
Across all three experiments, three patterns recur. First,
developer-test-pass is tightly constrained by PoC-pass: among the 734
case-level outcomes in our condition-by-language analysis
(aggregating all controlled conditions in
\S\ref{sec:major-experiments}), only a single case has
\codeword{PoC=fail} and \codeword{Dev=pass} simultaneously. Second,
the ratio of developer-test-pass to PoC-pass varies substantially by
condition and model, indicating that some benchmark configurations
improve oracle alignment much more than they improve true patch
quality. Third, the pool of “PoC-only” patches, i.e. cases where the
agent suppresses the PoC but fails the developer's test, remains
large across model generations: 38 cases for Sonnet 4.0, 41 for
Sonnet 4.5, and 33 for Opus 4.7. This persistent slab is exactly the
region that PoC-only benchmarks miscount as progress.

\section{Case Studies of Incidental Factors}
\label{sec:case-studies}

The experiments in \S\ref{sec:major-experiments} show how systematic
factors manifest at scale.
In this section, we walk through five concrete incidents drawn from
real runs with Claude Code paired with Claude Sonnet 4.5, in which a
single framework- or dataset-level decision either leaked the answer
to the agent or caused the pipeline to mis-score a correct patch.
We call these factors \emph{incidental} because they affect only a
subset of cases rather than the dataset as a whole.
Our goal is to show that for model training, leakage detection, or
leaderboard reporting, even rare incidents matter.
Every excerpt below is taken from raw agent traces. These traces are
archived in the artifact.

\subsection{Case 1: WebSearch Leakage}
\label{sec:case:websearch}

\noindent \textbf{Setup.}
This case study is observed in commit \codeword{69c6093} of the Go
package \codeword{xz}~\cite{xz_commit_69c6093},
where \codeword{ReadUvarint} can be made to consume unbounded input,
under a default Claude Code configuration with \codeword{WebSearch}
and \codeword{WebFetch} enabled.
This is the configuration that a benchmark operator gets ``out of the
box'' if they do not edit \codeword{.claude/settings.json}; many of
the other coding agents share this similar default configuration.

\noindent \textbf{Trace.}
The agent's first four tool calls after reading the bug description
were all web fetches surrounding \codeword{ReadUvarint}, with the
final call fetching the upstream patch commit directly.
Eleven turns later, the agent committed a fix and reported success
with the line ``The fix follows the same approach used in the Go
standard library fix,'' citing the very URL it had fetched.
Both the PoC and the developer's test passed.

\noindent \textbf{Mechanism.}
Public GitHub issues and security advisories are indexed by search
engines within hours of being filed.
For any benchmark derived from historical bugs, the ground-truth
commit is at most one web query away, and the case identifier or the
bug description usually contains enough hints (CVE id, GitHub issue
number, function name) to land the agent on it on the first try.

\noindent \textbf{Implication.}
For benchmarks targeting historical bugs, web tools must be disabled
by default and explicitly re-enabled only when the benchmark's claim
is about tool-use rather than about patching.
Our harness disables \codeword{WebSearch} and \codeword{WebFetch}
through the Claude Code \codeword{permissions.deny} list; frameworks
without an equivalent knob must isolate the container network.
Note that disabling these two tools is not in itself sufficient: an
agent that retains \codeword{Bash} can run \codeword{curl} or
\codeword{git fetch} against the public origin remote.
We address that channel in Case~2.

\subsection{Case 2: Git History Leakage}
\label{sec:case:gitlog}

\noindent \textbf{Setup.}
This case study is observed in commit \codeword{524f142} of the Go
package \codeword{bluemonday}~\cite{bluemonday_commit_524f142},
which is an XSS sanitizer regression where
\codeword{strings.ToLower()} is used to compare HTML tag names,
against a variant of our harness in which the project's full upstream
\codeword{.git} history was preserved.
This is meant to simulate the most common ``benchmark from a tarball
or unsanitized clone'' scenario.
The bug description shipped with the case mentions, as a hint, that
the issue is a regression of an earlier issue~\#56.

\noindent \textbf{Trace.}
The agent's reasoning before any code edit acknowledged the
regression hint provided by the bug description and decided to look
for the original fix from git history. It then issued a single
\codeword{git log} query, filtering by issue~\#56, and inspected the
resulting commit. By doing so, it found commit \codeword{5793ebc},
which contains a unified diff that has the exact API change as the
upstream fix for the current case.
The agent's subsequent edits track that diff almost line-for-line; it
concludes by quoting ``the fix pattern used before'' in its final message.
The case passes the PoC and the developer's test.

It is worth noting that the agent's invocation of \codeword{git log}
here is not malicious: it is a perfectly reasonable strategy for
understanding a regression, and a developer asked to do the same task
would do exactly the same thing.
The leakage is structural: any benchmark that ships the upstream
history alongside the bug effectively answers ``how was this fixed?''
in the workspace itself.

\noindent \textbf{Mechanism.}
The trigger commit is almost always followed, within a small number
of commits, by the upstream fix; for regression-style bugs, an even
older commit on a different file may already encode the answer.
A shallow clone, a tarball, or an explicit \codeword{git log}
sanitization step closes this channel.
Note that \codeword{WebSearch} (Case~1) and \codeword{git log} are
two routes to the same destination, and disabling either one alone is
insufficient.

\noindent \textbf{Implication.}
Every patching benchmark should document, and preferably automate,
its source-tree sanitization step.
In our default harness, the per-case Dockerfile clones the upstream
repository, checks out the trigger commit, then removes the git
history and initializes a new one containing only the buggy commit,
leaving the agent with a single-commit working tree.
The variant we used to produce this trace exists exclusively as a
control for the case study and is not part of the headline dataset.

\subsection{Case 3: Permission Misconfiguration}
\label{sec:case:permissions}

\noindent \textbf{Setup.}
In an earlier version of our harness, the agent ran as
\codeword{root} inside the case container, as a consequence, the
entire metadata directory was both readable and writable to it.
The motivation was operational: we wanted to avoid permission errors
during builds and dependency installation.
Rather than inlining build/test/PoC content into the prompt, we
instructed the agent to consult the metadata directory as needed.
The directory, however, also contains sensitive files used by the
post-agent evaluator, such as the upstream fix and the developer's test.
We re-created this configuration to drive the case in GitHub issue
GH-13931~\cite{php_issue_13931}
of the PHP project, which contains an ASan-detected null-pointer
dereference in \codeword{phpdbg}'s compiler shutdown path.

\noindent \textbf{Trace.}
The agent's first six tool calls all touched the metadata directory.
Listing this directory revealed the existence of the YAML config
file. By inspecting this file, the agent could see the path to the
ground-truth patch file used by the validator, which contains the
upstream patch verbatim.
Worse, the config file itself disclosed the upstream patch commit
hash (\codeword{c3acfb1b...}) and the issue URL, enough metadata to
pivot to Case~1's WebFetch leak if the tool is enabled.
The agent's subsequent edits to the code base are textually identical
to the upstream patch.

\noindent \textbf{Mechanism.}
The harness conflated two distinct sets of files: agent-visible task
inputs (config, bug report, build/test scripts, PoCs) and
evaluator-visible ground-truth metadata (upstream patch, developer's
tests, patch-commit references).
Bundling both into a single mount made permission misconfiguration
trivial: the agent did not need root, only \emph{read} access to the
wrong directory.

\noindent \textbf{Implication.}
The agent and the evaluator must run in disjoint mounts, and the
agent container should run as a non-root user.
The launch script materializes only the inputs the agent is allowed
to see (config without patch references, scripts, PoCs) into a
separate path before starting the agent; the metadata directory in
its full form is only mounted in the evaluator container.
We recommend a discipline of \emph{one write role, one read role}:
the agent can only read what the harness has explicitly staged for
it, and the evaluator's view is invisible to the agent.

\subsection{Case 4: Hidden Sub-Model Dependency}
\label{sec:case:haiku}

\noindent \textbf{Setup.}
This case is an outlier in the section: the failure mode does not
surface in the agent's transcript at all, and we therefore cannot
point to a log excerpt.
We discovered it indirectly, by cross-referencing per-case wall-clock
with model-side traffic in the LiteLLM admin panel.

\noindent \textbf{Symptom.}
Across many cases we observed unexpectedly long wall-clock times that
did not match any obvious build or test latency.
A given case might issue tens of \codeword{Bash} tool calls and
accumulate tens of minutes of overhead even when each individual
command, run outside the harness, completed in well under a second.
Re-running the same commands from a shell inside the same container
gave the expected sub-second timings, so the latency was not in the
project under test.

\noindent \textbf{Mechanism.}
Inspecting LiteLLM's request log revealed two distinct streams of
model traffic from a single agent run.
The first stream targets the model alias we had configured
(\codeword{anthropic/sonnet-4-5}) and corresponds to the main agent
loop; this stream succeeds normally.
The second stream targets the bare model name
\codeword{claude-haiku-4-5-20251001}, which our LiteLLM configuration
does not route, and corresponds to a per-Bash-command ``safety''
classification call that Claude Code issues internally.
Each such call hangs for $\approx\!120$ seconds---the underlying HTTP
timeout---before Claude Code falls back to allowing the command anyway.
The bash command then runs in a few milliseconds.
Schematically, every bash invocation looks like:
\textbf{(1)} Main model proposes \codeword{Bash(...)};
\textbf{(2)} Claude Code calls
\codeword{claude-haiku-4-5-20251001} for safety classification;
\textbf{(3)} no route; the call times out after $\approx\!120$\,s;
\textbf{(4)} Claude Code defaults to ``allow''; and \textbf{(5)} the
Bash command is finally executed.

\noindent The behavior is not visible in the agent's stream-JSON
output, in the harness logs, or in the LiteLLM logs of the configured
alias---only in the LiteLLM logs of the unrouted exact model name,
which an operator would have no reason to look at.
The behavior also appears to be hard-coded into the most recent
versions of Claude Code at the time we conducted our experiments: it
cannot be disabled through agent settings or prompt instructions.

\noindent \textbf{Workaround.}
We mitigate the issue by setting the lesser-known environment variable
\codeword{ANTHROPIC\_DEFAULT\_HAIKU\_MODEL}~\cite{claude_code_model_config}
in the agent
container, which redirects the safety-check slot to our routed alias.
With this set, per-Bash overhead drops back to near-negligible
latency and overall case wall-clock becomes consistent with the
underlying project's build and test latency.

\noindent \textbf{Implication.}
The effect of this issue is not random; it disproportionately
penalizes large projects in which the agent must navigate through
many directories or run many tests, exactly the cases where the
per-Bash $120$\,s tax compounds.
A benchmark operator who is not aware of raw model traffic will read
these as model-capability failures.
More broadly, every closed-source agent harness may call sub-models
the operator did not configure, or generally any hidden mechanics;
benchmark designs must be aware of the possibility that such
mechanics exist, monitor unexpected behaviors, and treat them as a run-invalidation event rather than a scored failure.

\subsection{Case 5: Agent-Written Test Collision}
\label{sec:case:testinterference}

\noindent \textbf{Setup.}
This case study is observed in commit \codeword{b8c9ede} of the
\codeword{Fiber} web framework~\cite{gofiber_commit_b8c9ede}
written in Go.
The upstream fix adds a test to \codeword{ctx\_test.go}, in addition to fixing the vulnerability.

\noindent \textbf{Trace.}
The agent produced a correct root-cause fix, matching the intention
of the upstream change.
Then it proactively rewrote the existing
\codeword{Test\_Ctx\_IsFromLocal} test in \codeword{ctx\_test.go} to
exercise the new code path. When doing so, it also added
\codeword{"net"} to the import block of \codeword{ctx\_test.go} so
the new code compiles.

\noindent \textbf{The collision.}
The upstream developer's test also modifies \codeword{ctx\_test.go}:
it renames \codeword{Test\_Ctx\_IsFromLocal} to
\codeword{Test\_Ctx\_IsFromLocal\_X\_Forwarded} and rewrites the body
to assert the same property the agent's added cases assert, and it
adds the same \codeword{"net"} import.
When the evaluator applies the developer's test using \codeword{git
apply} on top of the agent's commit:
the import-block hunk fails because the agent already added
\codeword{"net"}, while the function-body hunk fails because the
original \codeword{Test\_Ctx\_IsFromLocal} body that the diff expects
to find has been replaced by the agent's expanded version.
\codeword{git apply} aborts the entire diff, the developer's-test
phase is recorded as a failure, and the case is scored
\codeword{fail} despite the agent's patch being functionally
equivalent to (and arguably better tested than) the upstream one.

\noindent \textbf{Mechanism.}
Developer-test application is a textual diff, not a semantic merge;
any benign collision aborts the application.
The collision in this case is not adversarial---the agent did exactly
what its system prompt requested, namely add regression tests
covering the security property---and the failure lives entirely in
the scoring pipeline, not in the patch.

\noindent \textbf{Implication.}
Any added test files to the evaluation pipeline must be robust to
such benign collisions, or scoped to files the agent is unlikely to
touch. In our harness, before applying the added developer's test, we
revert all files that are modified in \codeword{test.diff} back to
their original states.
At minimum, the evaluator should distinguish \emph{test
addition/application failure} from \emph{test run failure}, reporting
the former as a pipeline error rather than as a patch
failure, otherwise correct patches are penalized whenever they
happen to touch the same test file the maintainer did, which for
security fixes is common.

\subsection{Why ``Incidental'' Factors Matter}
\label{sec:case:why-minor}

We refer to these factors as ``incidental'' because they do not
affect the majority of cases in full-dataset experiments and do not
consistently manifest in the same instances across repeated runs.
However, their impact is far from negligible. In our experiments,
each individual factor can appear in as low as $\approx5$\%
(Permission Misconfiguration) up to as high as $\approx32$\% (Test
Collision) of cases across one complete run on our full dataset, and
crucially, their effects are \emph{systematically biased} rather than
random. As a result, even a relatively small incidence rate can
distort aggregate metrics and downstream uses of benchmark data.

Despite their limited prevalence, incidental factors have a
disproportionate impact in two key contexts:

\noindent \textbf{Leaderboard reporting.}
Modern LLM-based patching systems are often compared using aggregate
success rates that differ by only a few percentage points. In this
regime, even low-frequency distortions can alter the relative ranking
of systems. For example, a framework issue affecting as little as
$5$\% of cases can shift the reported success rate enough to reverse
leaderboard positions\cite{swebenchleaderboard}. Because these
distortions are not independent of system behavior---for instance,
certain agents/models may be more likely to trigger timeouts, exploit
leakage channels, or generate conflicting tests---they introduce
\emph{systematic bias} rather than uniform noise.

This phenomenon mirrors concerns raised in prior work on benchmark
reliability and evaluation sensitivity, where small perturbations in
evaluation setup can lead to significant changes in reported
performance \cite{martinez_whats_2026, wang_are_2025}. In the context
of LLM agents, the problem is worsened by the complexity of the
evaluation pipeline, where infrastructure-level artifacts can
dominate model-level differences.

\noindent \textbf{Training data generation.}
The impact of incidental factors is even more pronounced when
benchmark traces are reused as training data for supervised
fine-tuning (SFT) or reinforcement learning (RL). In such pipelines,
each evaluated instance is treated as a labeled example, and
mislabeling directly affects the learned policy.

A case where a correct fix is incorrectly marked as failure---due to
timeout, infrastructure error, or test collision---becomes a
\emph{false negative}, discouraging behaviors that are in fact
desirable. Conversely, a case solved via unintended shortcuts, such
as WebSearch leakage or access to git history, becomes a \emph{false
positive}, reinforcing spurious strategies such as copying
ground-truth patches rather than reasoning about the bug.

This asymmetry is particularly problematic because learning
algorithms amplify consistent biases in the training signal.
Recently, studies of LLM benchmarking have highlighted how
contamination and evaluation artifacts can be absorbed into model
behavior during training \cite{xu_benchmark_2024, wang_are_2025}.
Incidental factors act as an additional source of such bias: although
they occur less frequently than systematic factors, their
per-instance impact on the training signal is disproportionately large.

\noindent
Taken together, these observations suggest that incidental factors
should not be treated as edge cases or experimental noise. Instead,
they represent a distinct class of evaluation failure modes that can
meaningfully affect both reported performance and the quality of
downstream models. A reliable benchmark must therefore detect,
report, and where possible eliminate these factors, rather than
averaging over them.

\section{Survey of Real-World Benchmarks}
\label{sec:survey}

The controlled experiments in \S\ref{sec:major-experiments} and case
studies in \S\ref{sec:case-studies} demonstrate that agent-level,
framework-level, and dataset-level factors can materially affect
reported patch quality.
We now ask: \emph{how prevalent are these pitfalls in existing benchmarks?}
To answer that, we surveyed 13 benchmarks and datasets that are
commonly used in prior work for evaluating LLM-based vulnerability patching.
Because these benchmarks differ in whether they provide an agent
interface or execution infrastructure, we organize them into three
categories (Table~\ref{tab:survey}) that determine which factors from
\S\ref{sec:factors} apply:
\noindent \textbf{LLM-agent patching benchmarks} provide an agent
harness or are used in prior work with LLM agents. All factors
(A1--A2, F1--F3, D1--D4) apply. We also include the SWE-bench family
in this category because it is widely used with LLM agents despite
being agent-agnostic. Although its tasks focus on general software
engineering rather than security vulnerabilities, it involves bug
repair workflows where many of the pitfalls we analyze still arise.

\noindent \textbf{Automated vulnerability repair (AVR) benchmarks}
are designed for traditional program-repair tools, with build and test
infrastructure but no agent harness. Dataset-level factors (D1--D4)
apply directly; framework-level factors (F1--F3) become relevant if
the benchmark is repurposed for agent evaluation. Agent-level factors
(A1--A2) do not apply, as these benchmarks do not define an agent
interface.

\noindent \textbf{Vulnerability collections} are curated datasets of
vulnerability metadata without execution infrastructure. Only
dataset-level factors (D1--D4) are relevant.

\newcommand{\na}{\textcolor{gray}{--}}
\begin{table*}[t]
  \centering
  \caption{Survey of 13 benchmarks and datasets used for evaluating
    LLM-based vulnerability patching
  }
  \label{tab:survey}
  \renewcommand{\arraystretch}{1.15}
  \footnotesize
  \begin{tabular}{@{}lclll@{\;\;}ccc@{\;\;}cc@{}}
    \toprule
    & \multicolumn{4}{c}{\textbf{Dataset}} &
    \multicolumn{3}{c}{\textbf{Framework}} &
    \multicolumn{2}{c}{\textbf{Agent}} \\
    \cmidrule(lr){2-5} \cmidrule(lr){6-8} \cmidrule(lr){9-10}
    \textbf{Benchmark}
    & \textbf{D1} & \textbf{D2} & \textbf{D3} & \textbf{D4}
    & \textbf{F1} & \textbf{F2} & \textbf{F3}
    & \textbf{A1} & \textbf{A2} \\
    \midrule
    \multicolumn{10}{@{}l}{\emph{LLM-agent patching benchmarks}} \\[2pt]
    SWE-bench~\cite{jimenez_swe-bench_2024}
    & test suite & issue + hints & shallow; part.\ san. & Py
    & delegated$^a$ & \checkmark & \ding{55}$^c$
    & agent-dep. & not contr.$^a$ \\
    MultiSWE-bench~\cite{zan_multi-swe-bench_2025}
    & test suite & issue & full; unsan. & 6 lang.
    & delegated & \ding{55} & \ding{55}$^c$
    & not doc. & not contr. \\
    Agentless~\cite{chunqiu2024agentless}
    & test suite & issue & full; part.\ san. & Py
    & N/A$^b$ & \checkmark & \na
    & fixed & N/A \\
    SEC-bench~\cite{hwiwon2025sec}
    & PoC & multi-source & full; unsan. & C/C++
    & partial & \ding{55} & \na
    & not doc. & not contr. \\
    PatchAgent~\cite{yu_patchagent_2025}
    & PoC + regr. & sanitizer rpt & full; part.\ san. & C/C++
    & tool-restr.$^d$ & \checkmark & \ding{55}$^e$
    & \checkmark & restricted$^d$ \\[4pt]
    \multicolumn{10}{@{}l}{\emph{Automated vulnerability repair (AVR)
    benchmarks}} \\[2pt]
    ARVO~\cite{mei2024arvoatlasreproduciblevulnerabilities}
    & PoC & issue + PoC & full; unsan. & C/C++
    & \ding{55} & \ding{55} & \na
    & \na & \na \\
    Vul4C~\cite{li2025sokeffectiveautomatedvulnerability}
    & PoC + regr. & NVD + crash trace & full / tarball; unsan. & C/C++
    & \ding{55} & \ding{55} & \ding{55}$^e$
    & \na & \na \\
    Vul4J~\cite{quang2022vul4j}
    & PoV tests & metadata only & full; san. & Java
    & \ding{55} & \ding{55} & \ding{55}$^e$
    & \na & \na \\
    VJBench~\cite{vjbench2023}
    & regr.\ tests & CVE + CWE ID & full; unsan. & Java
    & \ding{55} & \ding{55} & \ding{55}$^e$
    & \na & \na \\
    ExtractFix~\cite{gao_beyond_2021}
    & PoC & PoC only & full; unsan. & C/C++
    & \ding{55} & \ding{55} & \na
    & \na & \na \\[4pt]
    \multicolumn{10}{@{}l}{\emph{Vulnerability collections}} \\[2pt]
    Big-Vul~\cite{jiahao2020a}
    & none & CVE summary & \na & C/C++
    & \na & \na & \na
    & \na & \na \\
    CVEfixes~\cite{guru2021cvefixes}
    & none & CVE desc + msg & \na & 27 lang.
    & \na & \na & \na
    & \na & \na \\
    VADER~\cite{liu2025vaderhumanevaluatedbenchmarkvulnerability}
    & none & expert report & \na & 15+ lang.
    & \na & \na & \na
    & \na & \na \\
    \bottomrule
  \end{tabular}
  \\[2pt]
  {\raggedright\scriptsize
    \textbf{D1}\,=oracle granularity;\;
    \textbf{D2}\,=bug description;\;
    \textbf{D3}\,=source provenance (clone depth; sanitization status);\;
    \textbf{D4}\,=languages;\;
    \textbf{F1}\,=permission model;\;
    \textbf{F2}\,=timeout documented;\;
    \textbf{F3}\,=test integrity;\;
    \textbf{A1}\,=build/test instructions;\;
    \textbf{A2}\,=internet control.\;
    \checkmark\,=present/addressed;\;
    \ding{55}\,=absent/unaddressed;\;
  \textcolor{gray}{--}\,=not applicable.\par}
  {\raggedright\scriptsize $^a$SWE-bench is agent-agnostic;
    permission model and internet control depend on which agent
    framework is used (e.g., SWE-agent provides its own container, but
    the benchmark itself does not enforce isolation).\; $^b$Agentless
    delegates build/test to the SWE-bench Docker harness; it does not
    provide standalone scripts or an agent shell.\; $^c$SWE-bench and
    MultiSWE-bench apply \texttt{test\_patch} via \texttt{git apply},
    which can conflict with agent-modified files
    (\S\ref{sec:case:testinterference}).\; $^d$PatchAgent restricts the
    agent to three tools with no shell or web access, but does not
    enforce this via containerized network isolation.\; $^e$Benchmark runs the project's
    existing test suite on the agent-modified tree; the agent may
  modify or delete test files to influence the oracle outcome.\par}
\end{table*}

\subsection{Dataset-Level Observations}

\noindent \textbf{D1, oracle granularity.} Many security benchmarks
rely on PoC-only oracles, which overestimate
correctness (\S\ref{sec:exp:models}).
SEC-bench, ARVO, and ExtractFix determine correctness solely by
whether a single exploit input triggers a sanitizer report, which may
allow superficial fixes that suppress crashes without addressing root cause.
PatchAgent and Vul4C supplement the PoC oracle with the project's
existing regression tests;
Vul4J provides source-level PoV tests extracted from the upstream fixing commit;
VJBench relies entirely on the project's existing test suite without
any PoC execution.
The SWE-bench family uses test-suite oracles.

\noindent \textbf{D2, PoC format.} No benchmark documents or controls the format in which PoCs are
presented, despite its measurable effect on reported quality
(\S\ref{sec:exp:poc}).
Among the five security benchmarks that supply PoCs
(SEC-bench, PatchAgent, ARVO, ExtractFix, Vul4C), all use binary inputs exclusively.
Vul4J and VJBench side-step the issue by relying on source-level test
cases rather than PoC inputs, but this is an oracle-design choice
(D1), not a deliberate PoC-format decision.

\noindent \textbf{D3, source provenance and contamination.} How the
target source code is provisioned determines what incidental
signals are exposed to the agent.
SEC-bench, MultiSWE-bench, ARVO, ExtractFix, Vul4C, and VJBench all
retain full, unsanitized git history, making the upstream fix commit
discoverable via \texttt{git log} (\S\ref{sec:case:gitlog}).
Only SWE-bench's Python instances perform multi-step sanitization
(remote removal, tag pruning, reflog expiry, aggressive GC); its
non-Python instances only remove the remote.
PatchAgent removes \texttt{.git} from the patch-application copy but
not from the source-reading copy that the agent's \texttt{viewcode}
tool accesses, leaving git history partially exposed.
Vulnerability collections (Big-Vul, CVEfixes, VADER) distribute
source code as standalone snippets, database entries, or file
archives rather than as repository checkouts, so git-history-based
contamination does not arise from their design.
However, researchers who repurpose these datasets for LLM-agent patch
evaluation must supply their own source-retrieval step; naively
cloning the upstream repository introduces the same contamination
risk described above.

\noindent \textbf{D4, language balance.} Every security benchmark with
execution infrastructure covers a
single language family: C/C++ (SEC-bench, PatchAgent, ARVO,
ExtractFix, Vul4C) or Java (Vul4J, VJBench).
Vulnerability collections such as CVEfixes (27 languages) and VADER
(15+ languages) offer broader language coverage but lack infrastructure.
Among all surveyed benchmarks, only MultiSWE-bench combines
multi-language support with execution infrastructure.

\subsection{Framework-Level Observations}

\noindent \textbf{F1, permission model and harness isolation.} No
surveyed benchmark enforces a complete permission model that
limits agent access at both the application and network layers.
PatchAgent restricts agent actions to a defined tool set
(\texttt{viewcode}, \texttt{locate}, \texttt{validate}) with no shell
or web access.
SWE-bench and MultiSWE-bench are agent-agnostic: they delegate
isolation entirely to the agent framework (e.g., SWE-agent provides
its own container, but the benchmarks themselves do not enforce boundaries).
SEC-bench provides a partial harness with tag-based information
separation but no documented permission model.
AVR benchmarks provide no agent isolation, which is appropriate for
their original scope but
insufficient if repurposed for agent evaluation.
When permission models are weak or absent, the source-provenance
issues described in D3 become exploitable: agents with shell access
can discover ground-truth fixes via \texttt{git log} or by reading
mounted metadata.

\noindent \textbf{F2, timeouts and hidden mechanics.} Only PatchAgent
and SWE-bench document timeout thresholds; no
benchmark monitors hidden sub-model calls.
Without documented timeouts, capability failures are
indistinguishable from budget exhaustion (\S\ref{sec:case:haiku}).

\noindent \textbf{F3, test-patch collision.} SWE-bench applies its
\codeword{test\_patch} via \codeword{git apply}
without handling merge conflicts against agent-modified files;
MultiSWE-bench follows the same pattern.
Benchmarks that run the project's pre-existing test suite on the
agent-modified tree (PatchAgent, Vul4C, Vul4J, VJBench) face a
complementary risk: the agent may modify or delete test files,
inadvertently or deliberately, to influence the oracle outcome.
Any evaluation that runs tests on agent-modified code should address this
collision risk (\S\ref{sec:case:testinterference}).

\subsection{Agent-Level Observations}

\noindent \textbf{A1, build and test instructions.} Agent benchmarks
generally provide build and test infrastructure, but
whether and how these instructions reach the agent varies: SWE-bench
and MultiSWE-bench are agent-agnostic and leave prompt content to the
agent framework; PatchAgent abstracts building and testing behind its
\texttt{validate} tool; SEC-bench places build scripts inside the
container without documenting what the agent is told.
No benchmark ablates the impact of build and test instructions on
reported scores.
Because different harness configurations can shift PoC-pass rates by
up to 20\,\% (\S\ref{sec:exp:buildtest}), two studies on the same
dataset can report substantially different numbers without either being wrong.

\noindent \textbf{A2, internet and web-search access.} PatchAgent and
Agentless both prevent the agent from reaching the
internet with restricted agent workflows, but neither enforces this
at the network level.
SWE-bench and MultiSWE-bench leave network policy entirely to the
agent framework.
SEC-bench containers use default bridge networking, which permits
outbound connections.
No surveyed benchmark enforces network-level isolation (e.g.,
\texttt{{-}{-}network=none} or firewall rules).
For benchmarks derived from historical bugs, web search provides a
direct path to the upstream fix (\S\ref{sec:case:websearch}).

\subsection{Summary}

Among the five LLM-agent benchmarks, none addresses all nine factors.
Every pitfall identified in
\S\ref{sec:major-experiments}--\S\ref{sec:case-studies} is present in at least one surveyed benchmark.
PatchAgent covers the most ground---restricted tool set (F1),
documented timeouts and iterative budget (F2), and regression tests
supplementing the PoC oracle (D1). But it still lacks network-level
isolation (A2) and full git-history sanitization (D3).
SWE-bench provides strong source sanitization for Python instances
(D3) and documented timeouts (F2), but delegates all agent-level and
permission concerns to external frameworks (F1, A1, A2).
SEC-bench and MultiSWE-bench leave the majority of factors
unaddressed or undocumented.

AVR benchmarks and vulnerability collections are increasingly
repurposed for LLM-agent evaluation despite not being designed for that setting.
Repurposing either category requires supplying agent harnesses,
oracles, and isolation---engineering choices that introduce the very
confounds characterized in
\S\ref{sec:major-experiments}--\S\ref{sec:case-studies}.
In particular, unsanitized git histories (D3), absent permission
models (F1), and unrestricted network access (A2) become immediate
risks once an autonomous agent operates inside these environments.

\section{Guidelines for Reliable Benchmark Design}
\label{sec:guidelines}

We translate the principles established across
\S\ref{sec:major-experiments}--\S\ref{sec:case-studies} into a
checklist that benchmark designers and model benchmark operators can
apply directly.

\subsection{Guidelines Derived from the Principles}

\noindent \textbf{G1 (from \S\ref{sec:exp:buildtest}). Disclose the
harness's information budget.} Publish the exact prompt including any
case-specific instructions
provided, the list of tools, and the default values used in headline numbers.

\noindent \textbf{G2 (from \S\ref{sec:exp:poc}). Separate capability
from knowledge.} If the benchmark supplies PoCs, also report the
number without PoCs;
if it supplies build scripts, also report the number without them.
Diff-style ablations are cheap and disambiguate which part of the
headline reflects which axis.

\noindent \textbf{G3 (from \S\ref{sec:exp:models}). Report
developer-test results alongside PoC results.} At minimum, a
benchmark based on historical bugs should report both a
\emph{crash-suppression} rate and a \emph{developer-intent} rate,
understood as a lower bound on real patch quality.
We advocate for developer's tests derived from upstream commits as a
cheap, high-signal default.

\noindent \textbf{G4 (from F1/D3). Quarantine the ground truth.} Never mount
ground-truth artifacts in the
agent's container.
Automate source-tree sanitization (\codeword{.git} removal,
build-cache cleanup) as part of the setup of the harness environment.

\noindent \textbf{G5 (from F2). Be aware of hidden mechanics in
closed-source agents.} Monitor agent logs and raw LLM request logs to
detect hidden
behaviors. For closed-source agents with hidden sub-models, benchmark
operators should monitor per-slot model availability accordingly.

\noindent \textbf{G6 (from F3). Use collision-tolerant added test
application or classify collisions as pipeline errors.} Additional
tests (whether it is PoC-related, regression, or
developer's test) application should tolerate benign conflicts with
agent-added files, and should distinguish pipeline errors from patch
errors in the scoring output.

\noindent \textbf{G7 (cross-cutting). Publish failure taxonomies, not
just pass rates.} Each failed case should be labeled along the axes
used in this paper:
wrong build, regression-test failure,
developer-test failure, pipeline error, timeout, etc.
Without this, cross-benchmark comparison is impossible.

\subsection{A Reporting Checklist}

We propose a minimal reporting checklist that LLM vulnerability
patching benchmarking reports should include:
\textbf{(1)} The exact agent prompt(s) and any case-specific
instructions included.
\textbf{(2)} List of tools exposed to the agent, including network scope.
\textbf{(3)} Source tree provenance and sanitization steps.
\textbf{(4)} Oracle definition: PoC-only, regression, developer-test,
or combination.
\textbf{(5)} Inspection and analysis of the failure case logs with the
failure taxonomy of G7.

\section{Discussion and Threats to Validity}
\label{sec:threats}

\noindent \textbf{Scope and external validity.} Our dataset is
intentionally small and highly curated: 112 cases
across C/C++, Go, and Rust, restricted to projects with reproducible
Dockerized builds, executable PoCs, and developer-added tests. This
design improves experimental control, but it also introduces
selection bias~\cite{liu_critical_2021}. In particular, our cases
likely overrepresent better-maintained projects and bugs with cleaner
reproduction and validation artifacts. We therefore do not claim that
the absolute pass rates reported here generalize to the full
population of real-world bugs or to benchmark suites in other
language ecosystems. Our goal is narrower: to identify and measure
evaluation factors whose effects are large enough to remain visible
even in a controlled dataset. For that purpose, controlled
comparisons within a fixed harness are more important than scale alone.

\noindent \textbf{Our dataset versus real-world benchmarks.} Our
approach to uncovering benchmark pitfalls is to systematically
inject factors that may influence outcomes into our own dataset via
configuration flags, and then survey real-world benchmarks to assess
whether these factors are present.
We argue that this methodology is preferable to directly running
individual benchmarks: it allows us to keep both the agent harness
and dataset fixed while isolating and controlling each factor in
separate experiments, rather than being constrained by the design and
limitations of existing benchmarks.

\noindent \textbf{Oracle validity.} Developer's tests are a stronger
oracle than PoC-only evaluation, but
they are not a perfect proxy for patch
acceptability~\cite{qi_strength_2014,smith_is_2015}. They capture the
maintainer's notion of correctness only to the extent that the
upstream fix commit adds or meaningfully revises tests, and some
upstream tests may themselves be incomplete or overly narrow. To
reduce this threat, we exclude cases whose fix commits do not add
meaningful test coverage, and we derive developer's tests directly
from upstream artifacts rather than authoring them ourselves. Even
so, passing a developer's test should be interpreted as evidence of
closer alignment with maintainer intent, not as a proof of absolute correctness.

\noindent \textbf{Agent, model, and run dependence.} Absolute
performance depends on the chosen agent/model pair and on
run-level
stochasticity~\cite{lu_fantastically_2022,zhao_calibrate_2021}. Our
controlled experiments use OpenCode with Claude Sonnet 4.0 for
tractability, while our case studies use Claude Code with Claude
Sonnet 4.5 to expose behaviors of a production closed-source agent.
As a result, the exact percentages in \S\ref{sec:major-experiments}
should not be read as universal estimates of model capability. Our
main claims instead concern differences across conditions with the
dataset and harness held fixed---for example, the effect of
build/test instructions, PoC representation, and oracle choice. These
comparative patterns are less sensitive to the specific model choice
than the absolute rates themselves. We likewise place more weight on
large and consistent shifts (e.g., the 20-point C/C++ swing in
\S\ref{sec:exp:buildtest}) than on small percentage-point differences
that may fall within single-run variance.

\noindent \textbf{Memorization, contamination, and leakage.} Because
our cases are historical bugs drawn from public repositories,
some fixes may have appeared in model training
data~\cite{xu_benchmark_2024,cao_concerned_2024}. This threat mainly
affects absolute success rates. However, it does not explain the core
findings of the paper, which are driven by changes in benchmark
configuration while holding the underlying model fixed. Our
experiments show that build/test instructions
(\S\ref{sec:exp:buildtest}), PoC framing (\S\ref{sec:exp:poc}), and
framework-level leakage or timeout behaviors
(\S\ref{sec:case-studies}) substantially change measured outcomes
without changing the target bug or the model. A pure memorization
account also does not explain the persistent gap between PoC-pass and
developer-test-pass rates (\S\ref{sec:exp:models}): if success were
dominated by exact recall of upstream fixes, one would expect
substantial alignment with developer-intent oracles. We
therefore view memorization as a possible contributor to headline
pass rates, but not as a sufficient explanation for the
benchmark-sensitivity effects we report.


\section{Related Work}
\label{sec:related}

\noindent \textbf{LLM-based patching benchmarks and agents.} Recent work has
popularized real-repository issue resolution as a benchmark for code
agents, most notably SWE-bench and its multilingual
extensions~\cite{jimenez_swe-bench_2024}, together with agent
frameworks such as SWE-agent and
Agentless~\cite{yang_swe-agent_2024,chunqiu2024agentless}. Our goal
is not to provide another benchmark catalog; instead,
\S\ref{sec:survey} surveys existing benchmarks specifically through
the lens of evaluation pitfalls such as oracle choice, source
leakage, and configurations.

\noindent \textbf{Vulnerability-focused repair.} A parallel line of
work studies security-oriented repair using CVEs,
advisories, exploit inputs, or sanitizer
traces~\cite{mei2024arvoatlasreproduciblevulnerabilities,hwiwon2025sec,yu_patchagent_2025,quang2022vul4j,vjbench2023}.
These efforts are closer to our setting than general issue-resolution
benchmarks, but most still evaluate fixes primarily through PoC
suppression or existing regression tests. Our work is complementary:
rather than proposing a new leaderboard, we focus on how benchmark
design choices change the measured reliability of LLM-generated
security patches.

\noindent \textbf{Evaluation methodology.} Our closest antecedents are
studies questioning whether benchmark
scores faithfully measure the intended capability. In APR, Liu et
al.~\cite{liu_critical_2021} critically review the evaluation
practices used to assess repair systems, and prior work has long
shown that test-suite-passing patches can be plausible but incorrect
or overfit to weak
oracles~\cite{qi_strength_2014,smith_is_2015,long_analysis_2016}. In
the LLM era, recent audits of SWE-bench-style evaluation similarly
show that apparent issue resolution can diverge from true
correctness~\cite{martinez_whats_2026,wang_are_2025}. More broadly,
benchmark contamination and evaluation artifacts have become a
central concern in LLM
assessment~\cite{xu_benchmark_2024,cao_concerned_2024}, and the
machine-learning community has repeatedly warned that rapid benchmark
progress can outpace empirical rigor~\cite{sculley_winners_2018}. Our
contribution specializes these concerns to agentic vulnerability
patching, where reliability depends not only on the dataset and
oracle, but also on harness permissions, source sanitization, hidden
agent mechanics, and post-patch evaluation pipelines.

\noindent \textbf{Classical program repair.} Automated program repair
predates LLMs~\cite{le_goues_genprog_2012},
and many of the evaluation pathologies we study have clear precedents
in that literature, especially the tension between test-suite
adequacy and semantic
correctness~\cite{qi_strength_2014,smith_is_2015,long_analysis_2016}.
Our developer's-test proposal inherits this lesson, but adapts it to
modern agentic patching settings in which the model can localize,
edit, execute tools, and exploit benchmark infrastructure.

\noindent \textbf{LLMs for SE more broadly.} Broad surveys of LLMs for software
engineering~\cite{fan_large_2023,hou_large_2024} cover tasks such as
code generation, testing, vulnerability detection, and repair. Our
focus is narrower: the integrity of the benchmark pipelines used to
evaluate LLM-based patch generation.

\section{Conclusion}
\label{sec:conclusion}

Benchmarks for LLM-based automated vulnerability patching have become
a de facto measure of progress, but we show that their headline
numbers can be misleading. Reported success depends not only on the
model, but also on prompt design, framework behavior, source
packaging, and oracle choice. In particular, passing a PoC often
overstates true repair quality: across our experiments, models
improved much faster at suppressing benchmark symptoms than at
producing patches that satisfy developer-intent tests.

Our contribution is methodological. We introduced a taxonomy of
benchmarking pitfalls, a controlled multi-language dataset and
harness, and a set of experiments and case studies that make these
pitfalls visible and measurable. The resulting guidelines and
reporting checklist are meant to help future work distinguish model
capability from the highly sensitive evaluation pipeline.

If benchmark scores are to demonstrate the capability of a new model,
support credible scientific claims, or provide a reliable training
signal for future patching systems, then the evaluation pipeline
itself must be treated as part of the contribution. For this task,
what a benchmark measures is at least as important as how high the score is.

\appendix


\bibliographystyle{plainurl}
\bibliography{refs}

\section{Infrastructure Details}
\label{app:infra}

This appendix summarizes the evaluation harness used in
\S\ref{sec:method:agents}. The full artifact contains per-language
base images, per-case container recipes, agent images, a LiteLLM
proxy, and scripts for launching cases and validating ground truth.

\subsection{Container structure and isolation}
\label{app:infra:container}

Each run is built in three layers: (1) a language-specific base image
with the required toolchain, (2) a case image that checks out the
upstream repository at the trigger commit and installs case-specific
dependencies, and (3) an agent image that installs the agent runtime
and launches the patching workflow under a non-root \codeword{model}
user. For case-study controls involving source leakage, the harness
can deliberately relax these defaults, but the headline configuration
uses the isolated setup described below.

Inside the container, the project source tree is exposed at
\codeword{/src/<project>} and is the only location the agent is
intended to modify. Per-case metadata is staged separately under
\codeword{/ssebench}. In the default configuration, the agent runs as
an unprivileged user and cannot read ground-truth artifacts such as
the upstream patch or developer’s test. The evaluator, which runs
after the agent terminates, retains full access to
\codeword{/ssebench}. This separation is what allows us to study
permission and leakage failures as controlled deviations rather than
as properties of the default harness.

\subsection{Agent configuration}
\label{app:infra:agent}

A small Python driver constructs the user prompt from a fixed
template and launches the agent with its working directory set to the
checked-out source tree. The prompt is parameterized so that we can
independently toggle the factors studied in \S\ref{sec:exp:buildtest}
and \S\ref{sec:exp:poc}, including whether build/test scripts are
inlined and whether the PoC is exposed to the agent. Model selection
is likewise parameterized through a LiteLLM endpoint, allowing the
same harness to be reused across providers and model families.

For Claude Code, the harness writes a per-case settings file that
restricts tool permissions, disables web-search tools in the default
configuration, and redirects internal sub-model slots to the
configured LiteLLM endpoint. For OpenCode, the harness constructs an
equivalent configuration against the same endpoint. Unless otherwise
stated, all runs use a 1800-second wall-clock budget and a 500-turn budget.

\subsection{LiteLLM, networking, and resource limits}
\label{app:infra:runtime}

All model traffic is routed through a LiteLLM proxy, which provides a
uniform API surface across providers and preserves raw request logs
for post hoc analysis. This indirection is important both for
controlled model substitution and for diagnosing hidden agent
behavior, such as the internal safety-model traffic discussed in
\S\ref{sec:case:haiku}.

Each case runs in a resource-limited Docker container attached only
to an internal bridge network that exposes the LiteLLM proxy but not
the public Internet. This enforces the default no-network policy for
agents: even if application-level web tools are re-enabled,
shell-based network access (e.g., \codeword{curl} or \codeword{git
fetch}) remains blocked unless the harness is intentionally reconfigured.

\subsection{Evaluation and ground-truth validation}
\label{app:infra:eval}

After the agent terminates, the evaluator executes the four-phase
pipeline described in \S\ref{sec:method:oracle}: rebuild the project,
run the PoC(s), run the pre-existing regression suite, and then apply
the developer’s test and re-run the tests. Each phase is
timeout-bounded, and failures in applying additional test patches are
recorded as evaluation-phase failures rather than evaluator crashes.

Dataset curation uses a separate ground-truth validator built from
the same case image but without any agent. It applies the upstream
developer patch to a fresh checkout and runs the same four-phase
pipeline. A case is included in the dataset only if the upstream
patch passes all phases, ensuring that the benchmark instances remain
internally consistent as scripts and metadata evolve.

\clearpage
\onecolumn
\section{Dataset Inventory}
\label{app:dataset}

Table~\ref{tab:dataset-inventory} lists every case in our dataset: 25
C/C++, 62 Go, and 25 Rust cases (112 total). The \emph{Case ID}
columns contain the target project's name along with the case
identifier. The \emph{Bug Source} column reports the primary upstream
identifier (GitHub issue or pull request, CVE, GHSA advisory, RUSTSEC
advisory, or commit hash when no tracker entry exists). The
\emph{PoC} column abbreviates the PoC representation used by the
harness: \textit{src} is a source-level harness compiled and executed
against the project, \textit{text} is a textual input file (e.g.,
XML, Vim script, WebAssembly text), and \textit{bin} is a binary blob
fed to a project binary. Trigger and patch commits are git commit IDs
in short form.

{\renewcommand{\arraystretch}{1.30}\footnotesize
  \begin{longtable}{@{}p{0.36\textwidth}llcll@{}}
    \caption{Dataset inventory}\label{tab:dataset-inventory}\\
    \toprule
    Case ID & Lang & Bug Source & PoC & Trigger & Patch \\
    \midrule
    \endfirsthead
    \multicolumn{6}{c}{\tablename~\thetable{} (continued)}\\
    \toprule
    Case ID & Lang & Bug Source & PoC & Trigger & Patch \\
    \midrule
    \endhead
    \midrule
    \multicolumn{6}{r}{\textit{Continued on next page}}\\
    \endfoot
    \bottomrule
    \endlastfoot
    \texttt{cpython-py\_pr\_124555} & C/C++ & issue \#120378 & bin &
    \texttt{7bd9dbf} & \texttt{c2ba931} \\
    \texttt{exiv2-exiv2\_gh\_1901} & C/C++ & issue \#1901 & text &
    \texttt{4d55cb6} & \texttt{395389a} \\
    \texttt{exiv2-exiv2\_gh\_1913} & C/C++ & issue \#1913 & bin &
    \texttt{04f4624} & \texttt{be29681} \\
    \texttt{exiv2-exiv2\_gh\_2007} & C/C++ & issue \#2007 & bin &
    \texttt{aebb3fb} & \texttt{65fad9d} \\
    \texttt{exiv2-exiv2\_gh\_789} & C/C++ & issue \#789 & bin &
    \texttt{f05d100} & \texttt{8cd95e2} \\
    \texttt{hdf5-hdf5\_gh\_3842} & C/C++ & issue \#3842 & src &
    \texttt{3ea21cc} & \texttt{af4c6c4} \\
    \texttt{hdf5-hdf5\_gh\_4656} & C/C++ & issue \#4656 & src &
    \texttt{6cf3389} & \texttt{9fd4fd0} \\
    \texttt{hdf5-hdf5\_gh\_4705} & C/C++ & issue \#4705 & bin &
    \texttt{b11f2d0} & \texttt{aea4c53} \\
    \texttt{jasper-jas\_gh\_383} & C/C++ & issue \#383 & bin &
    \texttt{29154a5} & \texttt{2d6b6e8} \\
    \texttt{jq-jq\_gh\_2825} & C/C++ & issue \#2825 & text &
    \texttt{f86566b} & \texttt{3fa10e8} \\
    \texttt{libxml2-xml\_commit\_0bcd05c} & C/C++ & commit 0090675 &
    text & \texttt{0090675} & \texttt{0bcd05c} \\
    \texttt{libxml2-xml\_commit\_a820dbe} & C/C++ & commit db07dd6 &
    text & \texttt{db07dd6} & \texttt{a820dbe} \\
    \texttt{libxml2-xml\_gh\_931} & C/C++ & commit 24d7e15 & src &
    \texttt{24d7e15} & \texttt{71e1e8a} \\
    \texttt{libxml2-xml\_gh\_932} & C/C++ & commit 069bcda & src &
    \texttt{069bcda} & \texttt{499bcb7} \\
    \texttt{libxml2-xml\_gh\_933} & C/C++ & commit 24d7e15 & src &
    \texttt{24d7e15} & \texttt{71e1e8a} \\
    \texttt{php-php\_gh\_13094} & C/C++ & issue \#13094 & bin &
    \texttt{71a1def} & \texttt{1d6f344} \\
    \texttt{php-php\_gh\_13931} & C/C++ & issue \#13931 & bin &
    \texttt{5f9b9c4} & \texttt{c3acfb1} \\
    \texttt{vim-vim\_gh\_14738} & C/C++ & issue \#14738 & text &
    \texttt{8c35c26} & \texttt{6779719} \\
    \texttt{vim-vim\_gh\_16606} & C/C++ & issue \#16606 & text &
    \texttt{97baa1c} & \texttt{c5654b8} \\
    \texttt{vim-vim\_gh\_17005} & C/C++ & issue \#17005 & text &
    \texttt{c6336ac} & \texttt{f13c856} \\
    \texttt{wabt-wabt\_gh\_1524} & C/C++ & issue \#1524 & bin &
    \texttt{40f6a8a} & \texttt{713bece} \\
    \texttt{wabt-wabt\_gh\_2398} & C/C++ & issue \#2398 & bin &
    \texttt{c6853eb} & \texttt{874caa4} \\
    \texttt{wabt-wabt\_gh\_989} & C/C++ & issue \#989 & text &
    \texttt{21a9de4} & \texttt{943164d} \\
    \texttt{wabt-wabt\_gh\_991} & C/C++ & issue \#991 & text &
    \texttt{943164d} & \texttt{fb7e866} \\
    \texttt{wireshark-wireshark\_commit\_129565f} & C/C++ & commit
    10e9024 & text & \texttt{10e9024} & \texttt{129565f} \\
    \midrule
    \texttt{altcha-lib-go-4a56107} & Go & GHSA-6GVQ-JCMP-8959 & src &
    \texttt{361dbe3} & \texttt{4a56107} \\
    \texttt{apiserver-4df268e} & Go & GHSA-833M-37F7-JQ55 & src &
    \texttt{c17b7df} & \texttt{4df268e} \\
    \texttt{archiver-09b5706} & Go & CVE-2018-25046 & src &
    \texttt{7291196} & \texttt{09b5706} \\
    \texttt{avro-b4a402f} & Go & GHSA-9X44-9PGQ-CF45 & src &
    \texttt{3abfe1e} & \texttt{b4a402f} \\
    \texttt{aws-iam-authenticator-029d1dc} & Go & issue \#472 & src &
    \texttt{0a72c12} & \texttt{029d1dc} \\
    \texttt{beats-dd3af18} & Go & PR \#47839 & src & \texttt{a042afd}
    & \texttt{dd3af18} \\
    \texttt{beego-64cf44d} & Go & issue \#4946 & src &
    \texttt{daf779f} & \texttt{64cf44d} \\
    \texttt{beego-939bb18} & Go & GHSA-2J42-H78H-Q4FG & src &
    \texttt{1f40a88} & \texttt{939bb18} \\
    \texttt{beego-d5df5e4} & Go & GHSA-28R6-JM5H-MRGG & src &
    \texttt{d7a918f} & \texttt{d5df5e4} \\
    \texttt{bluemonday-524f142} & Go & issue \#111 & src &
    \texttt{3cce251} & \texttt{524f142} \\
    \texttt{chi-6eb3588} & Go & commit 6eb3588 & src &
    \texttt{de0d16e} & \texttt{6eb3588} \\
    \texttt{consul-template-d6a6f4a} & Go & commit d6a6f4a & src &
    \texttt{a8022bc} & \texttt{d6a6f4a} \\
    \texttt{docconv-b19021a} & Go & commit b19021a & src &
    \texttt{44d6739} & \texttt{b19021a} \\
    \texttt{echo-0ac4d74} & Go & issue \#2259 & src &
    \texttt{d77e8c0} & \texttt{0ac4d74} \\
    \texttt{etcd-7d1cf64} & Go & GHSA-M332-53R6-2W93 & src &
    \texttt{91b1a91} & \texttt{7d1cf64} \\
    \texttt{exporter-toolkit-2528877} & Go & GHSA-7RG2-CXVP-9P7P &
    src & \texttt{48868f5} & \texttt{2528877} \\
    \texttt{expr-0d19441} & Go & GHSA-93MQ-9FFX-83M2 & src &
    \texttt{77e876b} & \texttt{0d19441} \\
    \texttt{fiber-b8c9ede} & Go & GHSA-3Q5P-3558-364F & src &
    \texttt{b8c9ede} & \texttt{b8c9ede} \\
    \texttt{gin-81ac7d5} & Go & PR \#3500 & src & \texttt{d07db17} &
    \texttt{81ac7d5} \\
    \texttt{gin-bfc8ca2} & Go & issue \#2473 & src & \texttt{f3de813}
    & \texttt{bfc8ca2} \\
    \texttt{git-lfs-0cffe93} & Go & GHSA-6PVW-G552-53C5 & src &
    \texttt{e735de5} & \texttt{0cffe93} \\
    \texttt{gjson-192-f0ee9eb} & Go & issue \#192 & src &
    \texttt{9f58baa} & \texttt{f0ee9eb} \\
    \texttt{gjson-196-bf4efcb} & Go & issue \#196 & src &
    \texttt{9f58baa} & \texttt{bf4efcb} \\
    \texttt{gnark-crypto-5660088} & Go & issue \#1483 & src &
    \texttt{1873045} & \texttt{5660088} \\
    \texttt{go-rpmutils-a64058c} & Go & commit a64058c & src &
    \texttt{e314ffa} & \texttt{a64058c} \\
    \texttt{go-slug-764785b} & Go & commit 28cafc5 & src &
    \texttt{dbc66eb} & \texttt{764785b} \\
    \texttt{go.uuid-75cca53} & Go & issue \#73 & src &
    \texttt{36e9d2e} & \texttt{75cca53} \\
    \texttt{goutils-869801f} & Go & GHSA-XG2H-WX96-XGXR & src &
    \texttt{f192353} & \texttt{869801f} \\
    \texttt{goyave-5836bff} & Go & issue \#3293 & src &
    \texttt{fc57b4c} & \texttt{5836bff} \\
    \texttt{http-swagger-b7d83e8} & Go & commit b7d83e8 & src &
    \texttt{25e73d2} & \texttt{b7d83e8} \\
    \texttt{interceptor-fa5b35e} & Go & issue \#3148 & src &
    \texttt{7914551} & \texttt{fa5b35e} \\
    \texttt{jose2go-a4584e9} & Go & issue \#31 & src &
    \texttt{c5d6a57} & \texttt{a4584e9} \\
    \texttt{jsonparser-188-91ac968} & Go & issue \#188 & src &
    \texttt{1a29609} & \texttt{91ac968} \\
    \texttt{jwt-7b1c1c0} & Go & GHSA-29WX-VH33-7X7R & src &
    \texttt{9358574} & \texttt{7b1c1c0} \\
    \texttt{jwt-go-ec0a89a} & Go & issue \#422 & src &
    \texttt{32acca3} & \texttt{ec0a89a} \\
    \texttt{jwx-0e8802c} & Go & GHSA-PVCR-V8J8-J5Q3 & src &
    \texttt{ea70886} & \texttt{0e8802c} \\
    \texttt{jwx-64f2a22} & Go & GHSA-7F9X-GW85-8GRF & src &
    \texttt{55000b3} & \texttt{64f2a22} \\
    \texttt{jwx-d43f2ce} & Go & GHSA-HJ3V-M684-V259 & src &
    \texttt{778ed27} & \texttt{d43f2ce} \\
    \texttt{kubernetes-dad0e93} & Go & issue \#101695 & src &
    \texttt{90f5048} & \texttt{dad0e93} \\
    \texttt{mapstructure-742921c} & Go & GHSA-2464-8J7C-4CJM & src &
    \texttt{5d1b937} & \texttt{742921c} \\
    \texttt{minio-c1a4949} & Go & issue \#21647 & src &
    \texttt{334c313} & \texttt{c1a4949} \\
    \texttt{nuclei-0da993a} & Go & GHSA-7H5P-MMPP-HGMM & src &
    \texttt{4d5eb9c} & \texttt{0da993a} \\
    \texttt{oauth2-proxy-780ae4f} & Go & GHSA-4MF2-F3WH-GVF2 & src &
    \texttt{c377466} & \texttt{780ae4f} \\
    \texttt{opa-064f616} & Go & issue \#4762 & src & \texttt{243f526}
    & \texttt{064f616} \\
    \texttt{phonenumbers-152-0479e35} & Go & issue \#148 & src &
    \texttt{ccd8c4b} & \texttt{0479e35} \\
    \texttt{protobuf-go-f01a588} & Go & GHSA-8R3F-844C-MC37 & src &
    \texttt{235ef28} & \texttt{f01a588} \\
    \texttt{proxyprotocol-5c4a101} & Go & issue \#1 & src &
    \texttt{18c1c89} & \texttt{5c4a101} \\
    \texttt{rekor-39bae3d} & Go & GHSA-273P-M2CW-6833 & src &
    \texttt{812e699} & \texttt{39bae3d} \\
    \texttt{rekor-cf42ace} & Go & GHSA-2H5H-59F5-C5X9 & src &
    \texttt{46ac0b2} & \texttt{cf42ace} \\
    \texttt{revel-d160ecb7} & Go & issue \#1424 & src &
    \texttt{60c3d7a} & \texttt{d160ecb} \\
    \texttt{saml-8e92368} & Go & GHSA-5MQJ-XC49-246P & src &
    \texttt{2aeb2ef} & \texttt{8e92368} \\
    \texttt{saml-b07b16c} & Go & GHSA-267V-3V32-G6Q5 & src &
    \texttt{5f6e5ab} & \texttt{b07b16c} \\
    \texttt{schema-cd59f2f} & Go & GHSA-3669-72X9-R9P3 & src &
    \texttt{180f71e} & \texttt{cd59f2f} \\
    \texttt{shoutrrr-6a27056} & Go & issue \#240 & src &
    \texttt{2c9378b} & \texttt{6a27056} \\
    \texttt{submariner-operator-b27a04c} & Go & issue \#3041 & src &
    \texttt{7c94448} & \texttt{b27a04c} \\
    \texttt{text-23ae387} & Go & issue \#39491 & src &
    \texttt{3a82255} & \texttt{23ae387} \\
    \texttt{woodpecker-8aa3e5e} & Go & issue \#3929 & src &
    \texttt{904d8da} & \texttt{8aa3e5e} \\
    \texttt{xz-CVE-2020-16845} & Go & issue \#35 & src &
    \texttt{067145b} & \texttt{69c6093} \\
    \texttt{yaml-53403b5} & Go & issue \#89535 & src &
    \texttt{1f64d61} & \texttt{53403b5} \\
    \texttt{yaml-8f96da9} & Go & issue \#666 & src & \texttt{8f96da9}
    & \texttt{539c8e7} \\
    \texttt{yaml-bb4e33b} & Go & PR \#375 & src & \texttt{7b8349a} &
    \texttt{bb4e33b} \\
    \texttt{yaml-f221b84} & Go & PR \#515 & src & \texttt{bb4e33b} &
    \texttt{f221b84} \\
    \midrule
    \texttt{afire-da7904c} & Rust & commit da7904c & src &
    \texttt{5644d94} & \texttt{da7904c} \\
    \texttt{ammonia-142-4b8426b} & Rust & GHSA-5325-XW5M-PHM3 & src &
    \texttt{ae3fb56} & \texttt{4b8426b} \\
    \texttt{branca-7da3274} & Rust & issue \#22 & src &
    \texttt{aeadd4d} & \texttt{7da3274} \\
    \texttt{clatter-b65ae6e} & Rust & GHSA-253Q-9Q78-63X4 & src &
    \texttt{0fb7a00} & \texttt{b65ae6e} \\
    \texttt{cocoon-1b63921} & Rust & issue \#22 & src &
    \texttt{680b68c} & \texttt{1b63921} \\
    \texttt{cors-4c32059} & Rust & issue \#170 & src &
    \texttt{8d33ca4} & \texttt{4c32059} \\
    \texttt{fuel-vm-9c97c2b} & Rust & PR \#941 & src &
    \texttt{d8fa4ae} & \texttt{9c97c2b} \\
    \texttt{nano-arena-rustsec20210031} & Rust & RUSTSEC-2021-0031 &
    src & \texttt{f5306c7} & \texttt{6b83f9d} \\
    \texttt{rust-smallvec-rustsec20190009} & Rust & RUSTSEC-2019-0009
    & src & \texttt{dfd78d1} & \texttt{4ba0d0f} \\
    \texttt{rust-smallvec-rustsec20210003} & Rust & RUSTSEC-2021-0003
    & src & \texttt{7852204} & \texttt{9998ba0} \\
    \texttt{stack\_dst-rs-rustsec20210033} & Rust & RUSTSEC-2021-0033
    & src & \texttt{807e9d4} & \texttt{2a4d538} \\
    \texttt{string-interner-rustsec20190023} & Rust &
    RUSTSEC-2019-0023 & src & \texttt{6a9246d} & \texttt{04731e1} \\
    \texttt{hpack-d669282} & Rust & issue \#11 & src &
    \texttt{e833eca} & \texttt{d669282} \\
    \texttt{mongo-rust-driver-b918cd6} & Rust & PR \#1453 & src &
    \texttt{007894c} & \texttt{b918cd6} \\
    \texttt{nanoid-10-a902277} & Rust & GHSA-9HC7-6W9R-WJ94 & src &
    \texttt{d8d2113} & \texttt{a902277} \\
    \texttt{reffers-rs-6dd7ca0} & Rust & issue \#7 & src &
    \texttt{34d9878} & \texttt{6dd7ca0} \\
    \texttt{rs-stellar-strkey-83adad0} & Rust & issue \#58 & src &
    \texttt{9d308ac} & \texttt{83adad0} \\
    \texttt{rust-ordered-float-c55cda3} & Rust & RUSTSEC-2020-0082 &
    src & \texttt{e8fe06a} & \texttt{c55cda3} \\
    \texttt{rust-phonenumber-b792151} & Rust & RUSTSEC-2024-0369 &
    src & \texttt{c0fdd69} & \texttt{b792151} \\
    \texttt{rust-protobuf-f06992f} & Rust & RUSTSEC-2024-0437 & src &
    \texttt{8d045d2} & \texttt{f06992f} \\
    \texttt{rustix-31fd98c} & Rust & GHSA-C827-HFW6-QWVM & src &
    \texttt{4bb2f4b} & \texttt{31fd98c} \\
    \texttt{serde-yaml-b93aff6} & Rust & RUSTSEC-2018-0005 & src &
    \texttt{49e1e6f} & \texttt{b93aff6} \\
    \texttt{stack\_dst-2a4d538} & Rust & issue \#5 & src &
    \texttt{807e9d4} & \texttt{2a4d538} \\
    \texttt{string-interner-d91dac0} & Rust & issue \#9 & src &
    \texttt{f5ffa9d} & \texttt{d91dac0} \\
    \texttt{webp-62b4706} & Rust & issue \#40 & src &
    \texttt{d523893} & \texttt{62b4706} \\
  \end{longtable}
}
\twocolumn

\end{document}